\documentclass[12pt]{article}

\usepackage{colortbl}
\usepackage{amsmath}
\usepackage{amssymb}
\usepackage{appendix}
\usepackage{subfigure}
\usepackage{empheq}
\usepackage{framed}
\usepackage[USenglish]{babel}
\usepackage{graphicx}
\usepackage[margin=1.2in]{geometry}
\usepackage{pdfpages}

\date{}

\begin{document}

\title{
A doubly-magic storage ring EDM measurement method
}
\author{
Richard Talman \\
Cornell University, Ithaca, U.S.A.
}

\date{10 December, 2018}

\maketitle

\clearpage

\begin{abstract}
This paper discusses ``doubly-magic trap'' operation of 
storage rings with superimposed electric and magnetic bending, allowing
\emph{spins in two beams to be frozen} (at the same time, if necessary),
and their application to electric dipole moment (EDM) measurement.
Especially novel is the possibility of simultaneous 
storage in the same ring of frozen spin beams of two different 
particle types. A few doubly-magic cases have been found:
One has an 86.62990502\,MeV frozen spin proton beam and a 30.09255159\,MeV 
frozen spin positron beam (with accuracies matching their known magnetic 
moments) counter-circulating in the same storage ring. 
(Assuming the positron EDM to be negligibly small) 
the positron beam can be used to null the worst source of systematic EDM 
error---namely, the existence of unintentional and unknown average radial 
magnetic field $<B_r>$ which, acting on the MDM, causes spurious background 
spin precession indistinguishable from foreground EDM-induced precession. 
The resulting measured proton minus positron EDM difference is then 
independent of $<B_r>$. This amounts to being a measurement of the proton EDM.

Most doubly-magic features can be tested in one or more ``small'' EDM 
prototype rings. One promising example is a doubly-magic proton-helion 
combination, which would measure the difference between helion (i.e. helium-3) 
and proton EDM's. This combination can be used in the near future for EDM 
measurement, for example in a 10\,m bending radius ring, using only already 
well-understood and proven technology. In the standard model both EDM's are 
negligibly small. Any measureably large difference between these EDM values 
would represent ``physics beyond the standard model''.
\end{abstract}

\section{Doubly-magic EDM measurement method}

\subsection{Introduction}
\paragraph{Major previous EDM advances.\ }
Comparably important EDM advances that have been made in the recent past can 
be listed: 
The storage ring ``frozen spin concept'' according to which, for a 
given particle type, there can be a kinetic energy for which the 
beam spins are ``frozen'' in a storage ring---for example 
always pointing along the line of flight, Farley et al.\cite{Farley};
The recognition of all-electric rings with ``magic'' frozen spin kinetic 
energies (14.5\,MeV for electrons, 233\,MeV for protons) as especially 
appropriate for EDM measurement, Semertzidis et al.\cite{BNL-2011};
The ``Koop spin wheel'' mechanism, in which a small radial magnetic field 
$B_r$ applied to an otherwise frozen spin beam causes the beam polarization 
to ``roll'' around a locally-radial axis\cite{KoopSpinWheel} 
(systematic precession around any axis other than this would cancel any 
accumulating EDM effect); Spin coherence times long enough for EDM-induced 
precession to be measureably large, Eversmann et al.\cite{Eversmann}; 
``Phase-locking'' the beam polarization, which allows the beam polarization 
to be precisely manipulated externally, Hempelmann et al.\cite{Hempelmann}.  

\paragraph{Koop spin wheel.\ }
By design, the only field components in the proposed ring would be the radial 
electric component $E_x$, and ideally-superimposed magnetic bending would be provided 
by a vertical magnetic field component $B_y$. There also needs to be a tuneable 
radial magnetic field $B_r\equiv B_x$, both to compensate any uninentional and 
unknown radial magnetic field and to control the roll-rate of the Koop spin
wheel.

For a ``Koop spin wheel'' rolling around the radial $x$-axis, notes 
by I. Koop\cite{Koop-Juelich} provide formulas for the
roll frequencies (expressed here in SI units, with $B\rho$ in T.m),  
\begin{equation}
\Omega_x^{\rm B_x} = -\frac{1}{B\rho}\,\frac{1+G}{\gamma}\,cB_x,
\quad\hbox{and}\quad
\Omega_x^{\rm EDM} =  -\eta\frac{1}{B\rho}\,\bigg(\frac{E_x}{c} + \beta B_y\bigg).
\label{eq:roll-rates} 
\end{equation}
$G$ is the anomalous magnetic moment, $\beta$, $\gamma$ are
relativistic factors. $\Omega_x^{\rm EDM}$ is the foreground, EDM-induced 
roll frequency. $\Omega_x^{\rm B_x}$ is a roll frequency around the same radial axis,
caused by a radially magnetic field $B_x$ acting on the MDM. 
$cB\rho=pc/(qe)\equiv pc/(Ze)$ is the standard accelerator 
physics specification of storage ring momentum. The factor $\eta$ expresses 
the electric dipole moment $d=\eta\mu$ in terms of the magnetic moment 
$\mu$ of the beam particles.

\paragraph{Proposed EDM measurement technique.\ }
The proposed EDM measurement technique starts by measuring and nulling 
\begin{equation}
\Omega_x^{\rm Koop} = \Omega_x^{\rm EDM} + \Omega_x^{\rm B_x} \longrightarrow 0
\label{eq:roll-rates.2} 
\end{equation}
for the spin wheel of a secondary beam. (Not yet attempting 
simultaneously-circulating beams) the secondary beam is then dumped and, 
\emph{with no change of ring conditions whatsoever}, the matching frozen spin primary beam 
is stored. Since the primary beam is subject to the same radial magnetic fields
as the secondary beam, its $\Omega_x^{\rm EDM}$ roll rate will then provide a 
direct measurement of the primary beam EDM $d$. 

Previously one will, of course, also have followed Koop in minimizing 
$\langle B_x\rangle$, by measuring the  differential vertical separation of 
the two beams, which is similarly proportional to $\langle B_x\rangle$. 

Though technically more challenging, in some cases, for better control of 
systematic errrors, by running on different RF harmonics, the two beams can
circulate concurrently, running on appropriately-different RF harmonic 
numbers to compensate for their different revolution periods, with
circumferences matched to parts per million.

\paragraph{Polarimetry assumptions.\ }
Ultimate EDM precision may depend on resonant polarimetry, probably based 
on the Stern-Gerlach 
interaction\cite{RT-TheEDMChallenge}\cite{RT-PSTP-2017}\cite{RT-RAST}.
Meanwhile, impressive beam polarization control has been achieved using 
polarimetry based on left-right scattering asymmetry of protons or deuterons 
from carbon\cite{Hempelmann}, and much more progress will undoubtedly be made 
with this method. Any prototype EDM ring to be built in the near future will 
need to rely initially on this form of scattering asymmetry polarimetry. 

\subsection{Orbit and spin tune calculation}
\paragraph{Terminology.\ }
Fields are ``cylindrical'' electric ${\bf E}=-E_0{\bf\hat x}r_0/r$ and, 
superimposed, uniform magnetic ${\bf B}=B_0{\bf\hat y}$.
The bend radius is $r_0>0$. 
Terminology is needed to specify the relation between electric and magnetic 
bending:
Cases in which both forces cause bending in the same sense will be called
``constructive'' or ``frugal'';  Cases in which the electric and magnetic
forces subtract will be referred to as ``destructive'' or ``extravagant''.
There is a reason for the ``frugal/extravagant'' terminology to be favored. 
Electric bending is notoriously weak (compared to magnetic bending) and
iron-free (required to eliminate hysteresis) magnetic bending is also 
notoriously weak. 
As a result an otherwise-satisfactory configuration can be
too ``extravagant'' to be experimentally feasible.

A design particle has mass $m>0$ and charge $qe$, with electron charge 
$e>0$ and $q=\pm 1$ (or some other integer). These values produce circular 
motion with radius $r_0>0$, and velocity ${\bf v}=v{\bf\hat z}$, where the motion
is CW (clockwise) for $v>0$ or CCW for $v<0$. With $0<\theta<2\pi$ being 
the cylindrical particle position coordinate, the angular velocity is 
$d\theta/dt=v/r_0$. 

To limit cases we consider only electrons (including positrons) 
protons, deuterons, tritons, and helions; 
that is e-, e+, p, d, t, and h. The circulation direction of the so-called 
``master beam'' (of whatever charge $q_1$) is assumed to be CW or, equivalently,
$p_1>0$. 
The secondary beam charge $q_2$ is allowed to have either 
sign, and either CW or CCW circulation direction.

\paragraph{Fractional bending coefficients $\eta_E$ and $\eta_M$.\  }
(In MKS units) $qeE_0$ and $qe\beta c B_0$ are commensurate forces, 
with the magnetic force relatively weakened by a factor $\beta=v/c$ 
because the magnetic Lorentz force is $qe{\bf v}\times{\bf B}$. 
Newton's formula for radius $r_0$ circular motion can be expressed using 
the total force per unit charge in the form
\begin{equation}
\frac{F_{\rm tot.}}{e} =\beta\,\frac{pc/e}{r_0} = qE_0 + q\beta cB_0,
\label{eq:BendFrac.1} 
\end{equation}
Coming from the cross-product Lorentz magnetic force, the term $q\beta cB_0$
is negative for backward-traveling orbits because the $\beta$ factor 
is negative. The ``master'' beam travels in the ``forward'', CW direction. 
For the secondary beam, the $\beta$ factor can have either sign.
For $q=1$ and $E_0=0$, formula~(\ref{eq:BendFrac.1}) reduces to the standard 
accelerator physics ``cB-rho=pc/e''. 
For $E_0\ne 0$ the formula incorporates the relative ``effectiveness'' of
$E_0/\beta$ and $cB_0$.

Fractional bending coefficients $\eta_E$ and $\eta_m$ are then defined by
\begin{equation}
\eta_E = \frac{r_0}{pc/e}\,\frac{E_0}{\beta},
\quad\hbox{and}\quad
\eta_M = \frac{r_0}{pc/e}\, cB_0,
\label{eq:BendFrac.2} 
\end{equation}
neither of which is necessarily positive. 
They satisfy $\eta_E + \eta_M = 1$.

\paragraph{Spin tune expressed in terms of $\eta_E$ and $\eta_E$.\ }
With $\alpha$ being the angle between the in-plane component of
beam polarization and the beam direction, the ``spin tune'' is defined to
be the variation rate per turn of $\alpha$, expressed as a fraction of 
$2\pi$. Spin tunes in purely electric or purely magnetic rings are given by
\begin{equation}
Q_E = \Big(G - \frac{1}{\gamma^2-1}\Big)\,\gamma\beta^2 
    = G\gamma - \frac{G+1}{\gamma},
\quad
Q_M = G\gamma,
\label{eq:BendFrac.7}
\end{equation} 
With superimposed fields, the spin tune can be expressed in terms of
the fractional bending coefficients,
\begin{equation}
Q_S = \frac{d\alpha}{d\theta} = Q_E\,\eta_E + Q_M\,\eta_M.
\label{eq:BendFrac.6} 
\end{equation}

\paragraph{The ``magic energy'' condition.\ }
Superimposed electric and magnetic bending permits beam spins to be frozen 
``frugally''; i.e. with a ring smaller than would be required for all-electric 
bending. 
The magic requiremment is for spin tune $Q_S$ to vanish;
$$Q_S = \eta_{_E}Q_E + (1-\eta_{_E})Q_M = 0.$$
Solving for $\eta_{_E}$,
\begin{equation}
\eta_{_E} = \frac{G}{G+1}\,\gamma^2, \quad
\eta_{_M} = 1 - \frac{G}{G+1}\,\gamma^2.
\label{eq:SpinPrecess.5}
\end{equation}
For example, with proton anomalous moment $G_p=1.7928474$, trying $\gamma=1.25$, 
we obtain $\eta_{_E} = 1.000$ which agrees with the known proton 
233\,Mev kinetic energy value in an all-electric ring. 
For protons in the non-relativistic limit, 
$\gamma\approx1$ and $\eta^{\rm NR}_E \approx2/3$.
The magic electric/magnetic field ratio is
\begin{equation}
\frac{E}{cB}
 = 
\frac{\beta\eta_E}{\eta_M}
 =
\frac{\beta G\gamma^2}{1 + G(1-\gamma^2)} 
 =
\frac{G\beta\gamma^2}{1-G\beta^2\gamma^2}.
\label{eq:SpinPrecess.5pp}
\end{equation}

\paragraph{Wien filter spin-tune adjustment.\ }
Superimposed electric and magnetic bending fields allow 
small correlated changes of $E$ and $B$ to alter the spin tune 
without affecting the orbit. Being uniformly-distributed, 
appropriately matched electric and magnetic field components 
added to pre-existing bend fields can act as a (mono-directional)
``global Wien filter'' that adjusts the spin tune without changing 
the closed orbit. Replacing the requirement that $\eta_E$ and 
$\eta_M$  sum to 1, we require $\Delta\eta_M=-\Delta\eta_E$, 
and obtain, using the same fractional bend formalism,
for a Wien filter of length $L_W$ the spin tune shift caused by a 
Wien filter of length-strength product  $EL_W$ is given by 
\begin{equation}
\Delta Q_S^W
 =
-\frac{1}{2\pi}\,\frac{1+G}{\beta^2\gamma^2}\,\frac{EL_W}{mc^2/e}.
\label{eq:BendFrac.6pp} 
\end{equation}
For ``global'' Wien filter action by the bends of the entire ring, 
$L_W$ is to be replaced by $2\pi r_0$. 

\subsection{``MDM comparator trap'' operation}
\paragraph{Dual beams in a single ring.\ }
This section digresses temporarily to describe
the functioning of dual beams in the same ring as a ``spin tune 
comparator trap''.
A ``trap'' is usually visualized as a ``table-top apparatus''. 
For this paper ``table-top radii'' of $10$, $20$, or $50$, meters
(or rather curved sectors of these radii, expanded by straight sections
of comparable length) are considered.

Gabrielse\cite{Gabrielse-eEDM} has (with excellent justification) 
boasted about the measurement of the electron magnetic moment 
(with 13 decimal point accuracy) as ``the standard model's greatest
triumph'', based on the combination of its measurement to such high
accuracy and on its agreement with theory to almost the same 
accuracy.  Though other magnetic moments are also known to high
accuracy, compared to the electron their accuracies are inferior by 
three orders of magnitude or more. One purpose for a
spin-tune-comparator trap would be to ``transfer'' some of the electron's 
precision to the measurement of other magnetic dipole moments (MDM's). 
For example, the proton's MDM could perhaps be determined to almost the 
current accuracy of the electron's.

Different (but not necessarily disjoint) co- or counter-circulating 
beam categories include different particle type, opposite sign, dual speed, 
and nearly pure-electric or pure-magnetic bending.
Cases in which the bending is nearly pure-electric are easily 
visualized. The magnetic bending ingredient can be treated perturbatively. 
This is especially practical for the 14.5\,MeV electron-electron 
and the 233\,Mev proton-proton counter-circulating combinations.

\begin{figure}[ht]
\centering
\includegraphics[scale=0.32]{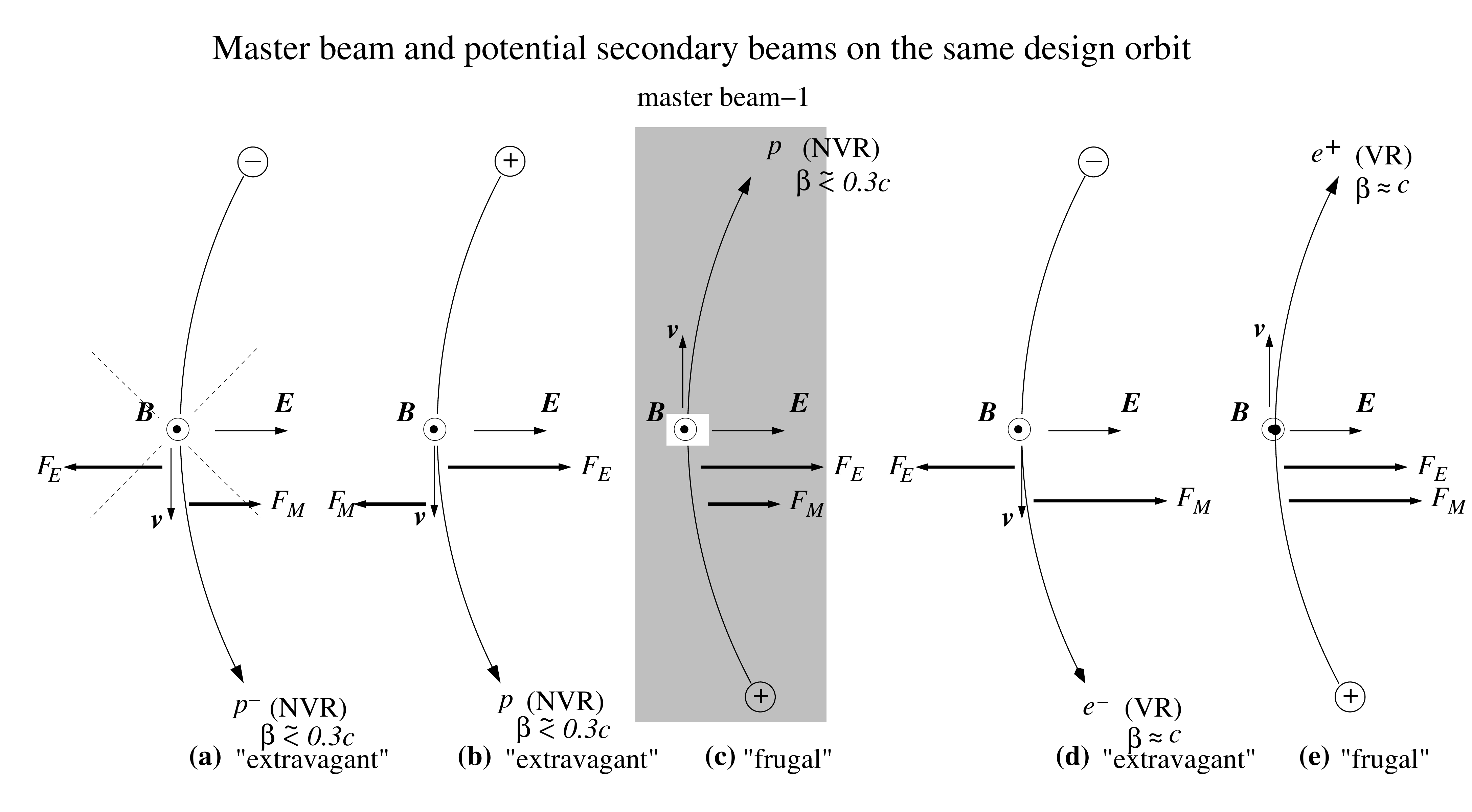}
\caption{\label{fig:Doubly-magic-cases}Examples of ``secondary beams''
designed to have the same design orbits as a (shaded) beam~1``master beam''.
Electric and magnetic force strengths are crudely represented by
the lengths of their (bold-face) vectors.
This figure is limited to very-relativistic (VR) electrons (of either sign)
and not-very-relativistic (NVR) protons (of either sign). CW and CCW orbits
are identical, except for traversal direction. For stable beam 
circulation the sum of electric and magnetic forces has to be centripetal. 
This condition is violated in case (a); the centrifugal electric
force exceeds the centripetal magnetic force.  }
\end{figure}

Storage of different beam types in the same ring is illustrated 
in Figure~\ref{fig:Doubly-magic-cases}. As explained in the 
caption, the bending can be either frugal or extravagant 
(i.e. constrictive or destructive). For a given particle type, if
the clockwise (CW) bending is frugal, the counter-clockwise (CCW)
bending is extravagant. For stable orbits the net radial force has
to be centripetal.  For the three cases described in this paper, 
the electric force magnitude exceeds the magnetic force magnitude.
This means that only positive particle beams can be stable. 

Eversmann et al.\cite{Eversmann} have demonstrated the capability
of measuring spin tunes with high accuracy. By measuring the spin
tunes of beams cirulating in the same ring (not necessarily simultaneously)
the MDM's of the two beams can be accurately compared. 

\paragraph{Sensitivity to imperfections.\ }
So far only perfect apparatus has been considered. Here we 
comment on imperfections. The main attribute to be claimed for the spin 
tune comparator will be its relative insensivity to imperfections. 
Whatever validity is claimed will come from a combination of 
(1) basing parameter determinations 
only on frequency measurement, (2) accurate knowledge of the MDM's, 
and (3) on the degree to which the spatial orbits of co- or counter-circulating 
beams are constrained to be identical to high accuracy. 
Also important
will be the degree to which the ratio of electric to magnetic field is constant 
around the ring.

(To be shown shortly) radial positioning errors are not a serious concern 
but requiring the design orbits to be accurately planar (i.e. lying in a 
single horizontal plane) markedly improves the MDM (and later the EDM) 
measurement accuracy.

The reason for controlling vertical orbit excursions to better
accuracy than horizontal has to do with spin precession control.
Let us assume that element positions are established initially to 
$\pm 100\,\mu$m accuracy horizontally, and $\pm10\,\mu$m accuracy vertically. 
Corresponding angular precision tolerances of about one-tenth milliradian horizontally 
and one-hundredth milliradian vertically will also be assumed.

Quoting G. Decker from 2005\cite{GlennDecker}
``Submicron beam stability is being
achieved routinely at many of these light sources in terms
of both AC (rms 0.1 - 200 Hz) and DC (one week drift)
motion.''
For fairly-smooth orbits, if the orbits are that close at all BPM 
locations, they will be almost that close everywhere.
With both spin tunes accurately measured, and their MDM's known,
the average circumference uncertainty will be dominated by spin 
tune measurement inaccuracy, which could correspond to 11 decimal
point circumference accuracy. 

In any case it is the circumference 
differences rather than the individual circumferences that will
govern the accuracy of the spin tune comparator.
After nulling all BPM differences, the CW and CCW circumferences
will then be equal to about 13 decimal places.

With revolution period known ``perfectly'' from RF frequency 
measurement, and average velocity known ``perfectly'' from frozen spin 
and accurately known MDM, even the absolute circumference value will 
be known to high accuracy.

\paragraph{Spin tune invariance and spin tune comparator trap precision}
By Eqs.~({\ref{eq:BendFrac.7}) 
spin tunes $Q_E$ and $Q_M$  depend only on $G$ and $\gamma$ but not
on bend radius $r_0$. This implies, \emph{for planar orbits,} that spin tunes are 
conserved constants of the motion, independent of horizontal steering 
errors---assuming, of course, that all ring components stay rigidly fixed 
in place, and with unchanged strengths.

But (because of commutativity failure for rotations around non-parallel axes) 
vertical steering errors prevent the spin tune formulas from being 
universally valid conservation laws.
Even so, from up-down symmetry, one expects the change $\Delta Q_S$ in
spin tune caused by a vertical deflection angle $\Delta y'$ to be proportional
to ${\Delta y'}^2$.  
By limiting the magnitudes of vertical deflection 
angles $\Delta y'$ to be less than, say $10^{-7}$, one can expect 
the spin tunes $Q_E$ and $Q_M$ to be independent of lattice errors to,
e.g. 14 decimal place accuracy.
Knowing the spin tunes and $\gamma$ values of both beams precisely,
and knowing the MDM of the particles in one of the beams, allows
the MDM of particles in the other beam to be determined to high accuracy. 

This is how a ``spin tune comparator trap'' can compare MDM's precisely.
Parameter tolerances for EDM measurement will be comparble to those
discussed in the previous section. 

\subsection{Secondary beam solutions}
\paragraph{Analytic formulation.\ }
Assume the parameters of a frozen spin master beam have 
already been established. As well as fixing the bend radius $r_0$,
this fixes the electric and magnetic bend field values $E_0$ and $B_0$. 
A further constraint that needs to be satisfied for 
secondary beam operation is implicit in the equations already derived.
To simplify the formulas we make some replacements and alterations, 
starting with
\begin{equation}
pc/e \rightarrow p,
\quad\hbox{and}\quad
m c^2/e\rightarrow m, 
\label{eq:Alterations.1}
\end{equation}
The mass parameter $m$ will be replaced later
by, $m_p$, $m_d$, $m_{\rm tritium}$, $m_e$, etc., as approppriate
for the particular particle type. These changes amount to switching the 
energy units from joules to electron volts and setting $c=1$. 

The number of ring and beam parameters can be reduced
by forming the combinations 
\begin{equation}
\mathcal{E} = qE_0r_0,
\quad\hbox{and}\quad
\mathcal{B} = qcB_0r_0.
\label{eq:Alterations.2}
\end{equation}
After these changes, the closed orbit condition has become  
\begin{equation}
p^4 -2\mathcal{B}p^3 + (\mathcal{B}^2-\mathcal{E}^2)p^2 - \mathcal{E}^2m^2=0,
\label{eq:AbbrevFieldStrengths.3}
\end{equation}
an equation to be solved for secondary beam momentum $p$. Any solution
meets the requirement for spin tune comparator functionality, but
not yet, in general, the doubly-magic, vanishing-spin-tune condition.

Any stable secondary beam orbit has to satisfy this equation but, because 
the electric and magnetic 
field values have been squared, not every solution of the equation 
has electric and magnetic field 
values that match the signs or magnitudes of the field values $E_0$ and $B_0$ 
constrained by the primary beam. So solutions of  
Eq.~(\ref{eq:AbbrevFieldStrengths.3}) have to be culled for consistency. 
The bending force has to be centripetal and consistent with bending in a 
circle of radius $r_0$.

By construction the already-established existence of a stable master beam implies
the existence of a real, CW (i.e. $p_1>0$) solution of the equation,
say with mass $m=m_1$. 
We look for other stable solutions, say with mass $m=m_2$ and momentum 
$p_2$, for which there are no parameter changes whatsoever, neither in $E_0$ 
nor $B_0$, nor in the sign or magnitude of the bend radius of curvature. 

For spin tune comparator functionality, satisfying 
Eq.~(\ref{eq:AbbrevFieldStrengths.3}) is sufficient for finding compatable dual 
beam parameters, including determining their spin tunes to the high precision 
with which the anomalous magnetic moments are known.

If anti-protons, anti-deuterons, or other anti-baryons were experimentally
available, the flexibility provided by Eq.~(\ref{eq:AbbrevFieldStrengths.3})
would be especially useful. The TCP combination of time, charge, and 
parity symmetry transformations would then provide TCP-matched solutions of 
the equation. But the only available negative particle is the negative electron, so
TCP invariance applies usefully only to beam combinations containing an electron or 
a positron beam. 

Limiting particle types to positron, proton, deuteron, tritium, and helions, 
a fairly comprehensive list of promising ``doubly-magic candidate'' solutions 
has been produced, satisfying these requirements, including the requirement 
that the master beam satisfy the magic beam condition. 

For EDM measurement functionality the further constraints to be met are severe. 
With parameters established and set such that the ``master beam'' is magic, 
the only remaining free parameter is the secondary beam energy. Doubly magic 
solutions are sought by varying this energy (always constraining the primary 
beam to satisfy the spin condition~\ref{eq:SpinPrecess.5}). As well as meeting 
the vanishing spin tune condition, the energy also has to be such that beam 
production and handling is practical, and high quality polarimetry is available.

\subsection{Three practical doubly-magic solutions}
\paragraph{Promising doubly-magic solutions.\ }
Several doubly magic beam pairs have been discovered. For this paper just 
three cases are considered. Their parameters are given in 
Table~\ref{tbl:Examples.1}. Details are given in the table caption and 
case by case explanations are given in the sequel.

Eq.~(\ref{eq:AbbrevFieldStrengths.3}) has been solved with MAPLE
to produce Table~\ref{tbl:Examples.1}. 
(Intended only for checking derived results, and otherwise unreliable)
the numerical anomalous magnetic moment values used have been:
\begin{align}
G\hbox{ [positron, e+]}\ &=\  0.00115965218076 \notag \\
G\hbox{    [proton, p]}\ &=\  1.79284735650 \notag \\
G\hbox{    [helion, h]}\ &=\ -4.18396274016
\end{align}

\hoffset -1.5cm
\begin{table}[h]
\small
\centering
\begin{tabular}{|cccccc||cccc|} \hline 
   r0   &  beam1  &    KE      &    E0   &     B0  & $\eta_E$ &   beam2   &   KE2   &    pc2  &    QS2    \\  
   m    &         &   GeV      &    V/m  &     T   &          &           &   GeV   &    GeV  &           \\ \hline 
\multicolumn{10}{|c|}{(b) PERTURBED DOUBLY-MAGIC PROTON-PROTON (nominal all-electric ring} \\ 
  50  &    CW p   &  0.2328  & 8.386e+06 & 1.6e-08 &       1  &   CCW p   &  0.2328   & -0.7007   & -2.144e-06   \\ 
         &          &         &         &         &          &    CW p   &  0.2328   &  0.7007   & -1.024e-15   \\  \hline
\multicolumn{10}{|c|}{(c1) DOUBLY-MAGIC PROTON-POSITRON (promising new option} \\  
  20  &    CW p   & 0.08663  & 6.355e+06 &   0.016 &   0.766  &  CCW e+   & 0.03009   & -0.0306   & 5.000e-06 \\ 
\multicolumn{10}{|c|}{(c2) DOUBLY-MAGIC POSITRON-PROTON (inverse of (c1))} \\  
  20  &   CW e+   & 0.03009  & 6.355e+06 &  -0.016 &   4.155  &   CCW p   & 0.08664   & -0.4124   & 5.842e-05 \\  \hline 
\multicolumn{10}{|c|}{ (q1) DOUBLY-MAGIC HELION-PROTON (JEDI currently-capable option)  } \\  
  10  &    CW h   & 0.03924  & 5.265e+06 &  -0.028 &   1.351  &   CCW p   & 0.03859   & -0.2719   & -6.173e-06  \\  
\multicolumn{10}{|c|}{ (q2) DOUBLY-MAGIC PROTON-HELION (inverse of (q1)) } \\  
  10  &    CW p   & 0.03859  & 5.265e+06 &   0.028 &  0.6958  &   CCW h   & 0.03924   & -0.4711   & 1.245e-05 \\  \hline 
\end{tabular}
\caption{\label{tbl:Examples.1}
Beam-pair combinations for the three EDM experiments discussed in this paper; master beam entries 
on the left, secondary beam on the right. ``(b)'', ``(c1)'', etc. are case labels.   
Dual rows allow either particle type to be designated ``master beam''. Candidate beam particle types 
are ''e+'',``p'', ``d'', ``t'', ``h'' labelling positron, proton, deuteron, triton, and helion rows. 
Bend radii, particle type, and kinetic energies are given 
in the first three columns. There is no fundamental dependence of spin tune $Q_s$ on $r_0$, 
but $r_0$ values have been chosen to limit $|E_0|$ to realistic values.
Bend radii choices of 10\,m, 20\,m, and 50\,m result from the compromise between reducing 
ring size and limiting electric field magnitude. $r_0$ can be increased beneficially 
except for cost in all cases, \emph{but not necessarily decreased}.
Master beam spin tunes are always exactly zero.
Spin tunes of secondary beams are given in the final column.
In all cases they are close enough to guarantee they can be tuned exactly to zero. 
Further, case by case, explanations are given in the text. 
}
\end{table}
\hoffset 1cm

Example (b) is perturbatively close to the already-known, singly-magic, all-electric 
solutions for protons. Examples (c1) and (c2) are doubly-magic solutions with positron and 
proton beams; the dually tabulated cases make the point that either beam can be interpreted 
as being the ``master beam''. Example (q1) and (q2) show doubly-magic solutions with proton and 
helion beams.

\paragraph{Perturbative variant of the nominal all-electric ring.\ } 
Case (b) in Table~\ref{tbl:Examples.1} represents the all-electric frozen-spin proton
ring which, up to now, has been implicitly anticipated to be the ultimate apparatus for
measuring the proton EDM. With its detailed features not yet understood this ring has
sometimes been called the ``holy grail'' ring.  Not intentionally pejorative, this language 
has been intended to acknowledge the significant uncertainties concerning the detailed 
properties of such a ring. In the table this ring is now referred to as the ``nominal 
all-electric ring''. 

In fact, case (b) is already a more realistic representation of the all-electric ring in the 
sense that some residual non-vanishing vertical magnetic field will be inevitable, 
even in an all-electric ring. This will require simultaneously-frozen-spin beam energies to have 
slightly different energies in all cases. 

With distributed electric and magnetic fields, using Eq.~(\ref{eq:BendFrac.6pp}) to 
describe the performance of the entire ring as a Wien filter, it will not be difficult to
meet the doubly-magic condition, even in the presence of extraneous weak vertical magnetic 
field. In itself, this would not justify distributed magnetic field, however, as the 
same trimming could be done with a short local Wien filter. 

However the ``perturbative'' solutions (available also for all-electric electron,
triton, and carbon\,13 frozen spin rings) are very robust in the sense that the 
superimposed magnetic field can be varied over a large range while preserving 
the doubly-magic capability. This opens up the possibility of investigating 
systematic EDM errors by varying the magnetic bending fraction by a large
factor.

This robust property applies uniquely to perturbations
away from an all-electric  ring. (In this case only) the structure of 
Eq.~(\ref{eq:AbbrevFieldStrengths.3})
guarantees that there is a continuum of doubly-magic solutions in the vicinity
of the all-electric condition. With counter-circulating beams of the same 
particle type, if the bending is frugal for one beam it is necessarily extravagant
for the other. But, since the sign of $\eta_M$ reverses at the all-electric point,
the continuity of solutions of Eq.~(\ref{eq:AbbrevFieldStrengths.3}) guarantees
the existence of a continuum of doubly-magic solutions in this vicinity.
This is the justification for attaching ``perturbed'' to the name of case (b).

There is a complication concerning RF frequency, in that slightly different beam
velocities will cause either slightly different orbits or slightly different
revolution periods. For slow particles, such as protons, this may 
require running on different harmonics of a single RF cavity. For positrons,
because they are fully relativistic, this would probably be impractical,
and the orbits would have to differ slightly. This RF issue is addressed
explicitly beow in the discussion of proton-helion case (q1).

\paragraph{Proton-positron doubly-magic solution.\ }
From the point of view of greatest promise for absolute EDM determinations, 
case (c1) (with equivalent case (c2)),
for proton and positron beams, seems to be the  most promising case. 
It enables measurement of the difference between a master beam
containing protons and a secondary beam containing positrons. 

Canceling the Koop wheel roll rate of the
secondary beam containing positrons cancels the 
radial magnetic field (under the assumption that the positron EDM is 
negligibly small). This  allows the primary beam Koop wheel 
roll rate to serve as a measurement of the proton EDM. 

As well as providing a clean, frequency difference measurement of the 
proton EDM, the beams can circulate simultaneously. 
Because positron and baryon velocities differ
by an order of magnitude, it is probably impractical
for the acceleration to be provided by harmonics of a single
RF cavity; dual RF systems will be needed.

A major impediment in this case is the low analysing power of
existing polarimetry methods for electrons (of either sign).
To remove the  ``holy grail'' qualification in this case will
require the development of resonant electron polarimeter. This
limitation is discussed further below. Achieving non-destructive,
high analysing power electron polarimetry seems likely to be
the only remaining major impediment to using EDM measurement
to test the ``standard model'' of particle physics.

\paragraph{Helion-proton solution, JEDI-capable option.\ }
From the point of view of earliest detection of physics beyond the 
standard model, case (q1) (with equivalent case (q2)) for proton and 
helion beams, seems the most promising.
Like the doubly-magic baryon-positron pair solutions, doubly-magic, 
different-type baryon-baryon pairs can be used to obtain EDM differences. 
A doubly-magic triton/proton solution has been found, but it requires 
electric fields that are probably unachievable, even in the 
largest ring currently under consideration.

However, by fortuitous accident of their anomalous magnetic moments, 
there is a doubly-magic helion/proton solution (q1) (with equivalent (q2)) 
that needs only a small ring. (The development of a polarized helion beam 
at BNL is described by Huang et al.\cite{BNL-helion}.) For this case
radius $r_0$ has been taken in the table, in round numbers, to be 10\,m.

The (q1) case has a CW, frugal bending solution for protons as master beam, 
with a CCW, extravagant bending helion beam as secondary beam. Carbon 
scattering asymmetry polarimetry will presumeably be used for both beams.

With a single RF cavity, to account for the different proton and helion 
velocities, the RF harmonic numbers can be 107 and 180, resulting in revolution 
period fractional difference of $3\times10^{-6}$. 

What makes this doubly-magic proton-helion option exciting is that, in the
near future, using only currently-established experimental techniques, an 
upper limit for the EDM of baryons can be substantially reduced from current
limits, possibly even to a level capable of demonstrating ``physics beyond 
the standard model''.  Some aspects of this ring are described in the
EDM prototype chapter of the present report, along with other applications 
of that proposed ring. 

\subsection{Gravitational effect EDM calibration}
Various 
authors\cite{Obukov-Silenko-1}\cite{Obukov-Silenko-2}\cite{Orlov-Flanagan}
have pointed out that general relativity (GR) introduces
effects that could be measureably large in proposed EDM rings. 
L\' aszl\'o and Zimbor\' as\cite{Laszlo-Zimboras}
calculate the GR influence on storage rings designed for
EDM measurement. 
The GR effect mimics the EDM effect.
Mistaken attribution to proton EDM produces a spurious proton EDM value 
of approximately $3\times10^{-28}$\,e-cm. 
This is about thirty times greater than the precision anticipated for
the all-electric, 233\,MeV ring originally contemplated and not inconsistent 
with an Orlov, Flanagan, Semertzidis\cite{Orlov-Flanagan} estimate. It is an 
accuracy that should be achievable with a small EDM prototype ring.

In a private communication, 
L\' aszl\'o and Zimbor\' as have kindly provided their result for the 
precession of the polarization vector $\vec{S}$; 
\begin{equation}
\frac{d\vec{S}}{dt} = \vec{\Omega}_{\rm GR} \times \vec{S},
\quad\hbox{where}\quad
\vec{\Omega}_{\rm GR}=G \gamma \vec{\beta} \times \vec{g}/c,
\label{eq:GR-EDM}
\end{equation}
where $\vec{\beta}$ is the velocity over speed of light of the particle,
$\gamma$ is its Lorentz factor, $G$ is its magnetic moment anomaly, $\vec{g}$ 
is the gravitational acceleration vector of the Earth, and $c$ is the speed 
of light, along with the interpretive comment ``It is seen that the pertinent 
gravitational contribution is a beam-radial precession vector, i.e. causes 
a contribution to the vertical polarizational buildup if the initial 
polarization was longitudinal.'' 

Comparing Eqs.~(\ref{eq:roll-rates.2}) and (\ref{eq:GR-EDM}) one notes
that the Koop roll and the GR roll depend differently on the parameters;
for example, in an all-electric ring, their dependence on velocity 
reversal is opposite.  This will help in subtracting the GR effect.
Once under control, the GR signal will serve as a valuable
calibrator of the EDM detection apparatus. The absolute level of this
calibration signal will be at the optimistic (i.e. large EDM value)
end of the range of plausible ``physics beyond the standard model''.

\subsection{The need for non-destructive resonant polarimetry}
Arthur Schawlow, co-inventor of the laser, is credited with the 
advice to ``Never measure anything but frequency''.  Though not emphasized up to
this point, this principle is implicit in the present paper. Though this
advice is often accepted, its basis is rarely explained. 

In our case the EDM signal at the end of an hour-long run may be an EDM-induced 
beam polarization angular difference of, say, a milliradian, between initial
and final beam polarization orientations.  Expressed as a fraction of a complete 
revolution of the beam polarization, this is $10^{-3}/(2\pi)$.
For any single run this angular shift
is likely to be comparable with the difference uncertainty of destructive polarimetry 
initial 
and final orientation measurements. (Then by averaging over, say, one thousand runs, 
the statistical error can be reduced by a factor of thirty or so. )

Consider the same hour-long run with non-destructive resonant polarimetry, assuming,
for the moment, the polarimeter natural resonant frequency to be the same as the beam 
revolution frequency. When sensed instantaneously, the resonator phasor angular 
advance from run beginning to
run end is likely to approximately match the $10^{-3}/(2\pi)$ difference of the previous 
paragraph, with ``phase noise'' having yielded approximately the same uncertainty. But
(absent other sources of low frequency noise) after non-destructive averaging the 
resonator phase for few-minute intervals at both beginning and end, the per-run 
phasor angular advance can be determined with far less uncertainty than is possible 
with destructive scattering asymmetry. 

This has not yet included two other factors that favor resonant 
polarimetry.  One of these factors is that the whole beam is measured at both
beginning and end. With destructive polarimetry, at best, orientation of only half 
of the beam is measured at run beginning; the other half of the beam is measured 
at the end.

The other advantage of resonant polarimetry would be that, in practice, the resonant 
polarimeter frequency will
be in the GHz range, 1000 times higher than the revolution frequency. Generally 
speaking, absolute precisions seem to increase inexorably as technological advances 
allow processing at ever higher frequaencies. But it would
not be legitimate to therefore claim a 1000 times higher precision, without having 
acquired a deeper understanding of the issues.   In our
case, for example, at every instant of time there will be a significantly large 
spread of particle revolution frequencies, more or less centered on a frequency
that is known  with exquisite accuracy from the known beam magnetic moments.  
Without having a clear understanding of the fluctuations and averaging it is
hard to refine the determination of the phase precision of resonant polarimetry. 

Regrettably, the entire discussion of resonant polarimetry up to this
point has been ``counting chickens before they're hatched''.  
Resonant polarimetry has never, in fact, been demonstrated to be
practical. However, theoretical calculations (admittedly due largely
to the present author)  based on the Stern-Gerlach
interaction, have shown that the regular passage of bunches of polarized 
electrons through a cavity should produce detectably-large cavity 
excitation\cite{RT-TheEDMChallenge}\cite{RT-PSTP-2017}\cite{RT-RAST}.
The latter two of these references describe, in considerable detail,
experiments being planned to test both transverse and longitudinal 
polarimetry, using a polarized electron linac beam in the CEBAF injection
line at the Jefferson Laboratory in Newport News, Virginia. Within
a few years tests like these should have resolved the issue conerning
the practicality of Stern-Gerlach polarimetry for electrons.

The proton's magnetic dipole moment is three orders of
magnitude smaller than the electron's. In the
absence of noise background a proton Stern-Gerlach signal reduced by
this factor, would still be detectably large but, without
extremely narrow band lock-in detection, the proton polarimetry signal is 
likely to be swamped by noise. This makes phase-locked-loop
proton beam polariization control based on resonant polarimetry likely
to fail, even if resonant electron polarimetry has been 
demonstrated to succeed. This is my expectation. 

It is this expectation that makes the doubly-magic proton-positron
combination for measuring baryon EDM's seem especially
important. With a positron beam phase-locked to resonant
Stern-Gerlach polarimeters (both transverse and longitudinal)
the Koop wheel manipulations, so optimistically assumed in the present 
paper, should, indeed be extremely precise for the positron beam. 

By exploiting the known relation between positron
and proton MDM's, it should then be possible to freeze the 
co-rotating proton spins just by controlling the positron beam 
spin tune and phase. With the frequency and phase of the proton beam 
magnetization then known to such high precision, the frequency 
filtering of a proton beam Stern-Gerlach resonator can be selective 
to reject the noise which would, otherwise, prevent the accurate
resonant determination of the magnetization signal. 

Only when non-destructive positron polarimetry has been successfully
demonstrated will it be legitimate to remove the ``holy grail'' designation
from the case (c1) positron-proton doubly-magic EDM ring design, to
make the discovery of physics beyond the standard model likely.
 
\subsection{The EDM measurement campaign}
The majority of my work in the storage ring EDM area for the last several 
years has been performed during, and in connection with, my stays at 
the IKP Institute for Nuclear Physics of Forschungszentrum, Juelich.

During 2018, in response to a CERN invitation, 
an EDM task force at the IKP laboratory has been performing a 
feasibility study of measuring electric dipole moments, especially of the proton. 
A full report is due by the end of the year.  The initial motivation for building a 
small prototype EDM ring was to demonstrate the ability to store enough 
protons to enable an EDM measurement in a storage ring with predominantly 
electric bending. A preliminary report was issued 
after the first quarter of 2018\cite{JEDI-Easter-report}.  
The present paper has been co-ordinated with this task 
force planning.

As well as developing long term planning, an important thrust of the task 
force has been to advocate the immediate development of designs for
a ``small'' EDM prototype storage ring. The doubly-magic design should 
have a major impact on motivation. This design eliminates the need to use the 
vertical separation of counter-revolving beam orbits to suppress radial 
magnetic field. Previous EDM designs have required excruciatingly small
vertical betatron tune in order to enhance this ``self-magnetometry'' 
sensitivity to vertical beam separation of counter-circulating beams. 
The coorrespondingly weak focusing was expected to set a small limit 
on the proton beam intensity.

The doubly-magic EDM ring design transfers this self-magnetometry resonsability
to a secodary frozen spin beam (with the admitted cost of measuring EDM differences
rather than absolute EDM values). Elimination of the need for ultraweak
focusing should enable the beam current intensities to be limited only by
previously-encountered understood effects.
This will permit the storage ring to have much stronger, alternating
gradient focusing, which can be expected to increase the achievable 
proton beam current substantially.

Another motivation for building a small prototype EDM ring has been
to develop and demonstrate the performance of instrumentation and procedures 
that will be 
needed for a subsequent larger ring. These applications are implicit in 
the examples of Table~\ref{tbl:Examples.1}. Especially relevant is
the doubly-magic combination of case (q1), which can be
used to measure the difference of proton and helion EDM's,
This can be done using carbon scattering polarimetry of the type that 
has been developed, and is already in service, in the Juelich COSY ring.
As already stated, any measureably large difference between proton
and helion EDM's would constitute physics beyond the standard model.

Important contributions by my EDM collaborators need to be acknowledged, 
especially to Sig Martin and Helmut Soltner for detailed discussions of implementation 
practicalities. Acknowledgements are also due to Maxime Perlstein for insisting 
on a less confusing treatment of the orbitry, to Eanna Flanagan and Andras Laszlo 
for communications concerning general relativistic effects, and to Andreas Wirzba 
for conveying and explaining a GR analysis by Kolya Nikolaev.

\end{document}